\documentclass[pra,twocolumn,nofootinbib,floatfix]{revtex4-1}
\usepackage[utf8]{inputenc}
\usepackage[T1]{fontenc}

\usepackage{graphicx}
\usepackage{amsmath, amsxtra, amssymb, mathtools, isomath, xspace}
\usepackage{bbold}

\usepackage{hhline}
\usepackage{multirow}
\usepackage{sidecap}

\long\def\symbolfootnote[#1]#2{\begingroup%
\def\thefootnote{\fnsymbol{footnote}}\footnotetext[#1]{#2}\endgroup}

\begin{document}

\title{Rapid production of large $^{7}$Li Bose-Einstein condensates using $D_1$ gray molasses}
\author{Kyungtae Kim}
\author{SeungJung Huh}
\author{Kiryang Kwon}
\author{Jae-yoon Choi$^{\ast}$}

\affiliation{Department of Physics, Korea Advanced Institute of Science and Technology, Daejeon 34141, Korea }
\symbolfootnote[1]{Electronic address: {jaeyoon.choi@kaist.ac.kr}}

\date{\today}

\begin{abstract}
 We demonstrate the production of large $^7$Li Bose-Einstein condensates in an optical dipole trap using $D_1$ gray molasses. The sub-Doppler cooling technique reduces the temperature of $4\times10^9$ atoms to $25~\mu{}$K in 3~ms. After microwave evaporation cooling in a magnetic quadrupole trap, we transfer the atoms to a crossed optical dipole trap, where we employ a magnetic Feshbach resonance on the $|F=1,m_F=1\rangle$ state. Fast evaporation cooling is achieved by tilting the optical potential using a magnetic field gradient on the top of the Feshbach field. Our setup produces pure condensates with $2.7\times10^6$ atoms in the optical potential for every 11~s. The trap tilt evaporation allows rapid thermal quench, and spontaneous vortices are observed in the condensates as a result of the Kibble-Zurek mechanism.
\end{abstract}

\maketitle
\section{Introduction}

Ultracold atoms have emerged as analog quantum simulators which can provide ideal platforms for studying quantum many-body problems~\cite{Bloch2012, Gross2017}. The Bose-Einstein condensation (BEC) of the $^7$Li atom is of particular interest because it is the lightest bosonic atom with a broad magnetic Feshbach resonance~\cite{chin2010}. Using the atoms, therefore, one can study correlated phases in the strongly interacting regime~\cite{Chevy2016} and develop a new form of quantum sensor composed of bright solitons that lack wave-packet dispersion~\cite{McDonald2014}. Moreover, the experimental compatibility with its fermionic ($^6$Li) isotope offers new chances to study the Bose-Fermi superfluid mixture~\cite{Ferrier2014}, and exotic ground states can be investigated in optical lattices~\cite{kuklov2003,Duan2003,Lewenstein2004}.

However, producing $^7$Li condensates is comparatively difficult compared with other alkali atoms because of two major limitations. First, the hyperfine structure of the $D_2$ excited state is not resolved so that a standard sub-Doppler cooling technique does not work efficiently. Second, it has poor scattering properties, and evaporation cooling works only in a limited parameter space window. For example, the upper hyperfine spin state $\left|F=2\right>$ has a negative-sign $s$-wave scattering length~\cite{Bradley1997}, and the collisional cross section shows a minimum at an energy of a few mK~\cite{Gross2008}. The lower hyperfine spin state $\left|F=1\right>$ has a very small scattering length under a residual magnetic field, so that evaporation cooling of laser-cooled atoms hardly works for both spin states in a conventional magnetic trap. As a result, the Bose-Einstein condensates are produced in an optical potential by sympathetic cooling with its fermionic isotope~\cite{Schreck2001,Ikemachi2017}, or using the Feshbach resonance~\cite{Gross2008} but with a very small numbers of atom.

These difficulties can be overcome by gray molasses on the $D_1$ transition line, which drops the temperature of atomic gases to a few recoil temperatures ($\sim$10~$\mu{}$K). The cooling technique has been demonstrated in various atomic species~\cite{Grynberg1994,Boiron1995,Esslinger1996,Boiron1996,Fernandes2012,Salomon2013,grier2013lambda,Burchianti2014,colzi2016sub} and, more recently, the condensation of $^7$Li atoms has been successfully observed by implementing the gray molasses~\cite{Dimitrova2017,Geiger2018}. In the experiments, the gray molasses offers an outstanding condition for evaporation cooling in a quadrupole magnetic trap, and BECs with atom number $N=1\sim4\times10^5$ have been generated in an optical potential after further evaporation cooling near a Feshbach resonance. 

Here, we elaborate the previous works and report the production of large $^7$Li condensates containing $N= 2.7\times10^6$ atoms with 11~s of duty cycle. The success of making large condensates lies in the efficient evaporation cooling in an optical trap by a trap-tilt evaporation scheme~\cite{hung2008}. It reduces the potential depth by tilting the optical potential without losing the trap confinement, contrasting conventional evaporation cooling by intensity ramp. The trap-tilt cooling technique allows a rapid thermal quench so that spontaneous vortices of the Kibble-Zurek mechanism~\cite{Kibble1976,Zurek1985} appear in the condensates. Besides, we observe that the gray molasses cooling can be further improved to reduce the temperature of the atoms captured in a magneto-optical trap (MOT) to 25~$\mu{}$K, which corresponds to 3.5 times the recoil temperature. We also present the evaporation path for each cooling stage, where nonadiabatic spin-flip atom losses at the magnetic quadrupole trap center are suppressed by a repulsive optical barrier~\cite{Davis1995}.

\section{Laser cooling}

\subsection{Magneto-optical trap}
Our experiment starts by collecting $^7$Li atoms in a magneto-optical trap from a Zeeman-slowed atomic flux. Three pairs of mutually orthogonal MOT beams are constructed by using two pairs of retroreflected light in the horizontal plane ($x$-$y$)  and one pair of counter propagating beams along the vertical $z$ direction. Each of the MOT beams contains both cooling and repumping light whose frequencies are $\Delta_{c,2}=-7~\Gamma$ and $\Delta_{r,2}=-4.7~\Gamma$, respectively [Fig.~1(a)], where  $\Gamma=2\pi\times5.87$~MHz is the natural linewidth of the excited state. The peak intensities of the laser beams are $I_{c,2}=3.3~I_s$ and $I_{r,2}=1.9~I_s$ ($I_s=2.54$~mW/cm$^2$ is the saturation intensity of the $D_2$ transition). An anti-Helmholtz pair of 42-turn water-cooled coils generates the magnetic quadrupole field, and we apply a field gradient of 20~G/cm along the axial $z$ direction in the MOT stage. After 5~s of loading time, we capture $6.4\times10^9$ of $^7$Li atoms in the MOT at a temperature of 1.6~mK. Then, the atoms are compressed by increasing the field gradient to 46~G/cm over 25 ms. In the last 2 ms of the compression, the frequency of the cooling (repumping) light is changed to $-1.5~\Gamma$ ($-15~\Gamma$), which reduces the beam intensity to 5$\%$ of the initial value at the same time. After the ramp, most of the atoms are cooled down to 900~$\mu$K.

\subsection{Gray molasses}
 
\begin{figure}
\includegraphics[width=0.9\columnwidth]{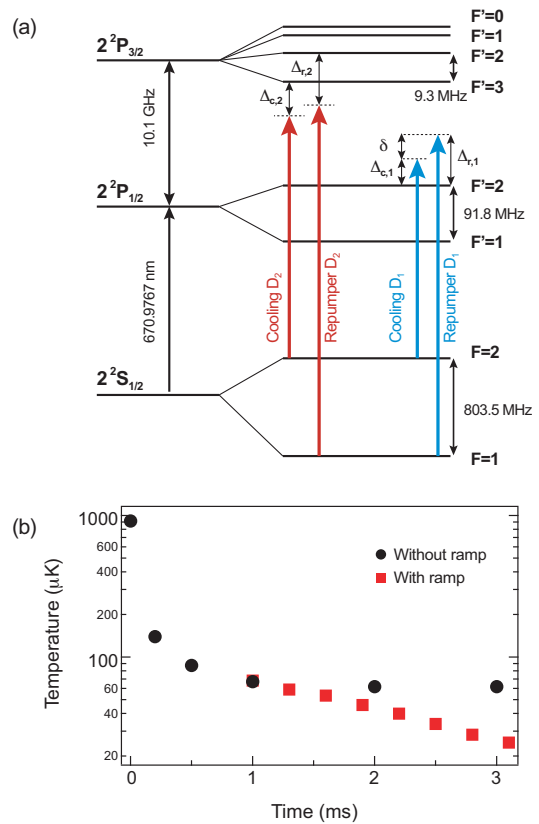}
\caption{The level structure of  $^7$Li atoms and laser cooling. (a) We use the $D_2$ transition line for the MOT [red (dark gray) arrow] and the $D_1$ transition line for the sub-Doppler cooling [blue (light gray) arrows]. (b) Temperature of the atoms during the molasses. The atoms in the compressed MOT are rapidly cooled down to $T\sim70~\mu$K in 1~ms and slowly settled to $\sim60~\mu$K after 2~ms (black circle). Changing the laser intensity and frequency, the temperature is reduced to 25~$\mu$K (red square). We also obtain a similar temperature by decreasing the molasses intensity to $I_{c1}=2.2~I_s$ at a constant frequency, but only 10$\%$ of the atoms remain. The temperature is measured by time-of-flight images at a various expansion time after pumping the atoms into the $\left|F=2\right>$ state using the MOT-repump beam. Each data point is averaged over 3--5 separate experimental runs, and the error bars are shorter than the marker size.  \label{LaserCooling}}
\end{figure}
The $D_1$ gray molasses consist of polarization gradient cooling and velocity-selective coherent population trapping~\cite{Aspect1988} in a Lambda-type three-level system and has been applied to lithium atoms~\cite{grier2013lambda,Burchianti2014}. In the report, $^6$Li gases are cooled down to 40~$\mu$K, serving as an essential step in the all-optical production of large degenerate Fermi gases~\cite{Burchianti2014}. Here, we employ the gray molasses to have a high collision rate in a magnetic trap, and thus generate large BECs after a rapid evaporation cooling.
 
The molasses beam is obtained from a high-power diode laser system using tapered amplifiers. The beam passes through a resonant electro-optical modulator (EOM) working at 803.5 MHz (hyperfine splitting frequency of the $^7$Li ground state), generating 2$~\%$ of the sideband for the repump light. For the molasses cooling, we set the laser frequency $\Delta_{r,1}=3.2~\Gamma$ and two-photon detuning $\delta=0$ [Fig.~1(a)]. Then, we superimpose the molasses light onto the MOT beams path, generating three orthogonal pairs of $\sigma^+-\sigma^-$ counter propagating beams. The $1/e^2$ beam waist is 5~mm at the trap center and each beam has a peak intensity of $22~I_s$. A pulse of 2~ms of gray molasses delivers 5.4$\times10^9$ number of the atoms in the compressed MOT at a temperature of 60~$\mu$K, which is similar to the previous experiments using lithium atoms~\cite{grier2013lambda,Burchianti2014}.

Like the gray molasses experiments using other alkali atoms~\cite{Fernandes2012,Salomon2013,colzi2016sub}, we are able to further cool $^7$Li gases by dynamically tuning the molasses beam parameters. After 1~ms of initial cooling at the maximal molasses lattice depth, the laser intensity and frequency are gradually changed to 12~$I_s$ and $\Delta_{r,1}=6.6~\Gamma$ with $\delta=0$, respectively. As shown in Fig.~1(b), we observe the temperature drops, and $4\times10^9$ atoms reach $25~\mu$K at the optimal conditions. The lowest temperature in the experiment corresponds to $3.5 ~T_{\text{rec}}$, where $T_{\text{rec}}=\hbar^2 k_\text{L}^2/m k_B$ is the recoil temperature ($\hbar$ is the Planck constant $h$ divided by 2$\pi$, $k_\text{L}$ is the cooling laser wave number, $m$ is the atomic mass, and $k_B$ is the Boltzmann constant). We also observe that the stray magnetic field $B_{\text{ext}}$ reduces the coherence of the dark state~\cite{Salomon2013}. The final temperature increases quadratically as a function of the external magnetic field, $\Delta{}T=B_{\text{ext}}^2\times84(8)~\mu$K/G$^2 $, so that a compensating residual magnetic field of less than 100~mG is necessary to reach few recoil temperatures.

\section{Evaporation cooling}
To generate large atom number condensates, we follow two-step evaporation cooling after the gray molasses: evaporation cooling is first taking place in a magnetic trap, and then the atoms are transferred to an optical dipole trap for further evaporation and Bose-Einstein condensation~\cite{Truscott2001,Dimitrova2017,Geiger2018}. This is attributed to the scattering properties of $^7$Li atoms, which has a negative scattering length  $a_0=-27.6~a_B$ ($a_B$ is the Bohr radius) in the upper hyperfine $\left|F=2\right>$ state so that the condensate atom number is limited to a few thousands~\cite{Abraham1997}. The lower hyperfine $\left|F=1\right>$ state, on the other hand, has a too small scattering length ($a_0=6.8~a_B$) for efficient evaporation~\cite{Ikemachi2017}, calling for a strong magnetic field ($\sim$700~G) to tune the scattering length~\cite{chin2010}. Therefore, large $^7$Li condensates can be produced with the $\left|F=1\right>$ state after evaporation cooling in an optical potential near the Feshbach resonance. However, an optical dipole trap has limited trap volume and potential depth compared to a magnetic trap, so that we first cool the atoms in the $\left|F=2,m_F=2\right>$ state in a magnetic potential and then transfer them to a crossed optical dipole trap. After 5~s of full evaporation cooling in the magnetic and the optical potential, we obtain a pure $^7$Li condensate with $2.7\times10^6$ atoms in the $\left|F=1,m_F=1\right>$ state.


\subsection{Magnetic quadrupole trap}
\begin{figure}
\includegraphics[width=0.9\columnwidth]{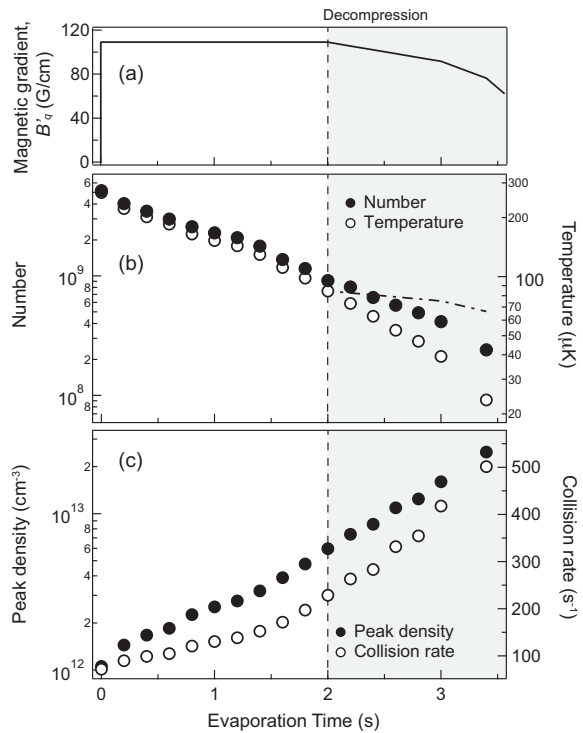}
\caption{Evaporation cooling in the magnetic quadrupole trap. (a) Magnetic field gradient, $B_q'$, during the evaporation process. Time evolution of (b) atom number, temperature, (c) peak density, and collision rate. (b) The dash-dotted line represents the estimated temperature solely from the adiabatic decompression without evaporation cooling. The atom number and temperature are determined from absorption images after $t_e\geq13$~ms of the time of flight. (c) We open the trap (gray zone) after 2~s of evaporation to keep the central peak density below $3\times10^{13}$cm$^{-3}$. The data points consist of  eight independent measurements and the error bars are smaller than the points size.  \label{QTevap}}
\end{figure}

After the gray molasses, evaporation cooling takes place in a magnetic quadrupole trap. The quadrupole trap is helpful for efficient evaporation because of its tight confinement and offers sufficiently large optical access to the cold atoms thanks to its simple coil geometry. In the experiment, we generate the quadrupole field using the same coil pairs in the MOT, and focus a blue-detuned 532~nm laser light at the trap center to suppress the Majorana atom loss~\cite{Davis1995}.  The laser beam propagates along the $x$ direction, and the effective potential for the $\left|F=2,m_F=2\right>$ spin stretched state becomes
 \begin{equation}
 U(\mathbf{r})=\mu_B B_q'\sqrt{\frac{x^2+y^2}{4}+z^2}+U_pe^{-2\left(\frac{y^2+z^2}{w^2}\right)}-mgz,
 \end{equation}
where $\mu_B$ is the Bohr magneton, $B_q'$ is the field gradient along the $z$ axis, and $g$ is the acceleration of gravity. The plug beam waist is $w=25~\mu$m, and 10~W of the laser beam generates a repulsive potential barrier height $U_p=k_B\times716~\mu$K at the zero-field center.

Before turning on the magnetic trap, we optically pump the atoms to the stretched state since most of the atoms after the gray molasses are in the $\left|F=1\right>$ state. The atoms are pumped via the $\left|F=2\right>\rightarrow\left|F'=2\right>$ $D_1$ transition using a laser light that contains two different frequencies, $\Delta_{c,1}=2.5~\Gamma$ and  $\Delta_{r,1}=5.8~\Gamma$. The pump beam travels in the horizontal plane in a retro-reflected configuration with circular polarization so that the $\left|F=2,m_F=2\right>$ state becomes a dark state to the pumping light. We shine the laser light for 150~$\mu$s under a 3~G of bias field, pumping almost all of the atoms into the stretched state. In order to load the atoms with high density, the field gradient is switched on to generate 58~G/cm in 100~$\mu$s by discharging a capacitor, and gradually increase to 109~G/cm in 0.2~s. 

Here, the frequency ramp is not implemented for additional cooling of the gray molasses to have a higher initial density in the magnetic trap. Without the frequency ramp, the optical pumping increases the temperature of atoms in the gray molasses to $70~\mu$K, and after the magnetic trap transfer, we have $N\simeq 5.1\times10^9$ and $T=270~\mu$K. When using the frequency ramp, on the other hand, the atoms are heated up to 40~$\mu$K by the dark state pumping, and then it rises to 240~$\mu$K in the quadrupole trap with $3.6\times10^9$ atoms. Still, the gray molasses provides a very favorable condition for evaporation cooling. Without the gray molasses, the initial temperature in the quadrupole trap is $\sim2$~mK, and we cannot achieve the runaway evaporation cooling because of the scattering length drops at a few mK~\cite{Gross2008}. 


Forced evaporation cooling is performed by applying a microwave frequency on the $\left|2,2\right>\rightarrow\left|1,1\right>$ hyperfine spin state transition. We linearly sweep the microwave frequency from 840 to 820~MHz in 2~s and to 811~MHz in another 1.4~s. To prevent strong atom losses due to dipolar relaxation and the three-body molecular recombination~\cite{Gerton1999}, the field gradient is reduced to 76~G/cm in the last 1.4~s of evaporation [Fig.~2(a)]. The time evolution of the atom number $N$ and temperature $T$ during the evaporation is displayed in Fig.~2(b), from which we estimate a peak density $n=N/[32\pi(k_BT/\mu_BB_q')^3]$ and elastic collision rate $\gamma=n\sigma\bar{v}$, respectively. Here, $\sigma=8\pi a_0^2$ is the elastic scattering cross section and $\bar{v}=(16k_BT/\pi m)^{1/2}$ is the mean relative velocity. The collision time, $\tau=1/\gamma$, decreases during the frequency sweep, demonstrating  runaway evaporation cooling in the quadrupole trap [Fig.~2(c)]. We observe the collision rate is $10^3$ higher than the loss rate of the trapped atoms, ensuring thermal equilibrium during the evaporation process.  
\begin{figure}
\includegraphics[width=0.9\columnwidth]{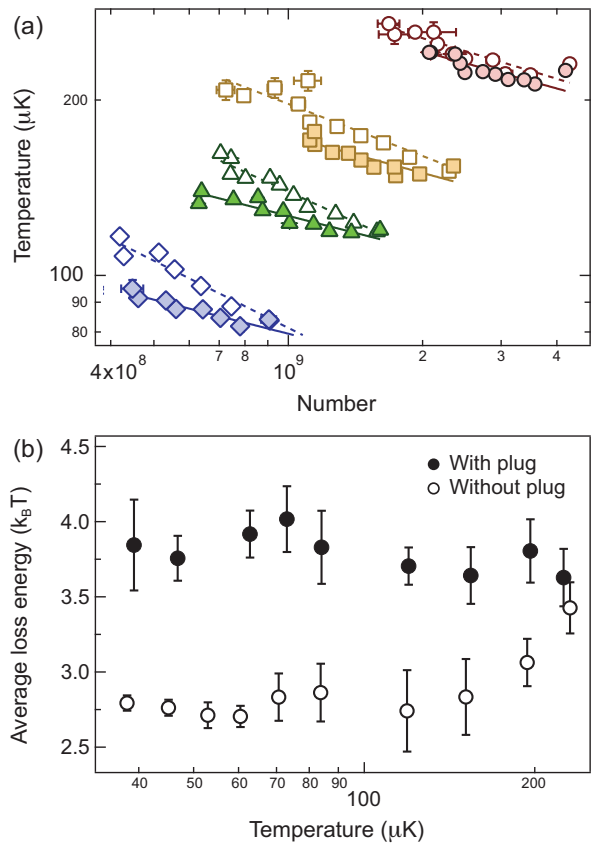}
\caption{Majorana heating and atom loss in the magnetic quadrupole trap. (a) Evolution of temperature by the atoms loss with (closed symbol) and without (open symbol) the optical plug potential under various initial temperature, $T(0)=230~\mu$K (red circle), $T(0)=150~\mu$K (yellow square), $T(0)=120~\mu$K (green triangle),  and $T(0)=84~\mu$K (blue diamond). The solid (dashed) lines represent the power-law fit curve with (without) the plug beam. (b) The average loss energy per atom loss vs initial temperature with (closed circle) and without (open circle) the plug beam. We take 3--5 measurements for each data point, and the error bars represent one standard deviation of the mean and the fit uncertainty.
\label{Majorana}}
\end{figure}

\subsection{Effects of optical plug beam}
The thermodynamics of atoms trapped in the quadrupole trap by the Majorana loss has been well characterized by the rate equations for atom number $N$ and temperature $T$~\cite{Petrich1995,Heo2011,dubessy2012rubidium},
\begin{eqnarray}
\frac{\dot{T}}{T}&=&-\left(\frac{\varepsilon_m}{\varepsilon}-1\right)\Gamma_m,\\
\frac{\dot{N}}{N}&=&-\Gamma_m-\Gamma_b.
\end{eqnarray}
Here, $\varepsilon_m$ is the mean energy per atom loss by the nonadiabatic spin flip, $\varepsilon=4.5~k_BT$ is the average energy of the atoms in the linear trap, and $\Gamma_b$ is the background loss rate. The $\Gamma_m$ is the Majorana loss rate~\cite{Petrich1995},
\begin{equation}
\Gamma_m=\chi\frac{\hbar}{m}\left(\frac{\mu_BB_q'}{k_BT}\right)^2,
\end{equation} where $\chi$ is a dimensionless geometrical factor, measured to be about 0.16~\cite{Heo2011,dubessy2012rubidium}. In the research~\cite{Heo2011}, the optical plug beam enhances the lifetime of the trapped atoms by reducing density at the trap center. The average energy per lost atoms $\varepsilon_m$, however, is not affected by the optical plug beam, implying that the atom loss still mostly occurs near the trap center.

In this section, we investigate the mean loss energy $\varepsilon_m$ in the quadrupole trap and observe the clear effect of the optical plug beam. The background loss rate in the quadrupole trap is measured to be $\Gamma_b=0.0093(8)$~s$^{-1}$ and can be neglected in the rate equations. Then, the dynamical evolution of the temperature can be expressed as a function of atom number, 
\begin{equation}
T(t)/T(0) = [N(t)/N(0)]^{(\varepsilon_m/\varepsilon-1)},
\end{equation}\label{NT}and we measure the $\varepsilon_m$ from the power-law exponent in the hold-time dynamics of $N$ and $T$. Figure~3(a) shows temperature dynamics at various initial conditions with and without the plug beam. The initial temperature $T(0)$ and atom number $N(0)$ are set by the microwave frequency, which is turned off during the hold time to exclude the evaporation effect. The atom loss leads to heating of the system, which becomes more evident at low temperature and without the optical plug beam.

These observations are reflected in the temperature dependence of the average loss energy as shown in Fig.~3(b). Without the plug potential, the $\varepsilon_m$ decreases as the temperature is reduced and saturates to 2.8(1)~$k_BT$. This can be explained by the Majorana heating process, which becomes the dominant heating source for a cold-atomic sample ($\propto T^{-2}$) and is expected to show 2.5~$k_BT$ of mean loss energy~\cite{dubessy2012rubidium}. The increases of the $\varepsilon_m$ at higher temperature can be attributed to a residual heating mechanism such as current noise in the magnetic trap and inelastic collisional events, rather than the Majorana loss. We speculate that an imperfection of polarization in the pumping beam induces the inelastic collisional loss at the beginning of the microwave evaporation. At sufficiently low temperature ($k_BT\ll U_p$), the spin-flip atom loss in the optically plugged quadrupole trap can cool the gases since the atoms have to climb up the plug hill ($\varepsilon_m>\varepsilon$). However, we observe that it stays around 3.9(1)~$k_BT$ without noticeable temperature dependence [Fig.~3(b)].

\subsection{Crossed optical dipole trap}
\begin{figure}
\includegraphics[width=0.9\columnwidth]{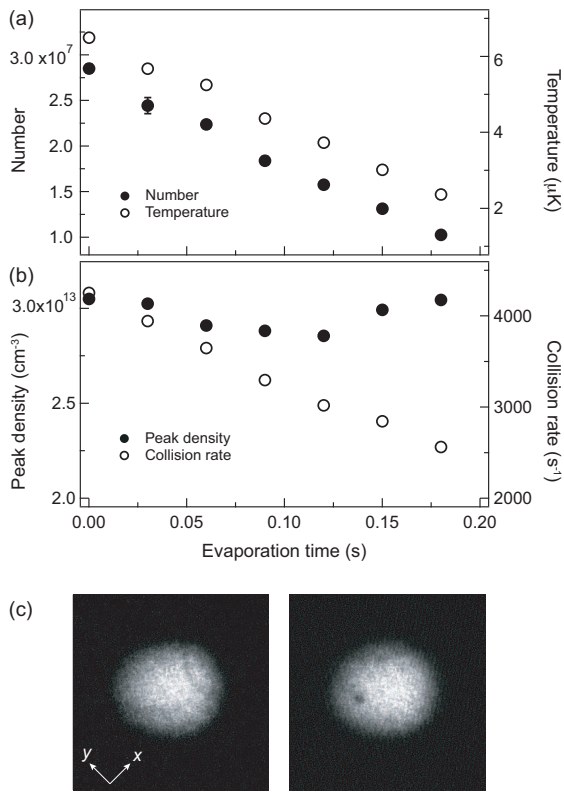}
\caption{Evaporation cooling for Bose-Einstein condensation in the crossed dipole trap. (a) Atom number, temperature, (b) peak density, and collision rate as a function of evaporation time. Data points represent mean values of five individual realizations, and the error bars denote the standard deviation of the mean. (c) Absorption images of BEC with a vortex. The condensates are obtained after rapid thermal quench (details described in the main text), and the vortex is represented as a density-depleted region in the condensates after 18~ms of expansion time. \label{CT}}
\end{figure}
As a final step to produce BECs, we load the cold atoms into a crossed optical dipole trap. The optical trap consists of two laser beams with 1064 and 1070~nm wavelength, propagating in the horizontal plane at a folding angle $\theta\simeq 90^{\circ}$. The laser beams are crossed at $300~\mu$m away from the quadrupole trap center so that the optical trap has a negligible influence on the optical suppression of the Majorana atom loss. At the crossing point, we have a beam radius of $156~\mu{}$m for 1064~nm light and $205~\mu{}$m for 1070~nm laser, respectively.  

After the microwave evaporation in the magnetic trap, the dipole potential is gradually turned on, producing $U_0=42~\mu{}$K of potential depth in 300~ms, and the field gradient is ramped down to zero in 600~ms. To maximize loading efficiency, we evaporate the atoms by applying a linear microwave frequency sweep from 811 to 804~MHz in 600~ms. About $5\times10^7$ number of atoms at $9~\mu{}$K are transferred in the crossed trap. Then, the atoms are prepared in the $\left|1,1\right>$ state by a Landau-Zener sweep. By turning on a uniform bias field of 700~G along the $z$ direction, the scattering length is set to about $100~a_B$. After the Feshbach field ramp, the magnetic field gradient $B_q'$ is turned on to 12~G/cm. This produces a linear potential in the $z$ axis that lowers the potential depth and cools the atoms without losing the trap confinement~\cite{hung2008}. 
 
By linearly increasing the field gradient to 30~G/cm in 300~ms, we achieve a rapid evaporation in the optical potential [Fig.~4(a)]. The trap beam intensity is simultaneously lowered to maintain the peak density, $n\simeq 3\times 10^{13}$~cm$^{-3}$. The truncation parameter $\eta=U_0/k_BT$ is increased from 5.5 to 7.5. After 180~ms of evaporation, a Bose-Einstein condensation is observed from the bimodal density distribution in a time-of-flight image [Fig.~5(a) inset]. The BEC transition temperature is  $T_c=2.4~\mu$K (calculated $T_c=2.9~\mu$K) with the critical atom number $N_c=10^7$. After an additional 120~ms of evaporation, a pure condensate of $2.7\times 10^{6}$ atoms is obtained. The trapping frequencies at the end of the evaporation are $(\omega_x,\omega_y,\omega_z)=2\pi\times$(165, 280, 324)~Hz. 

\begin{figure*}
\includegraphics[width=1.9\columnwidth]{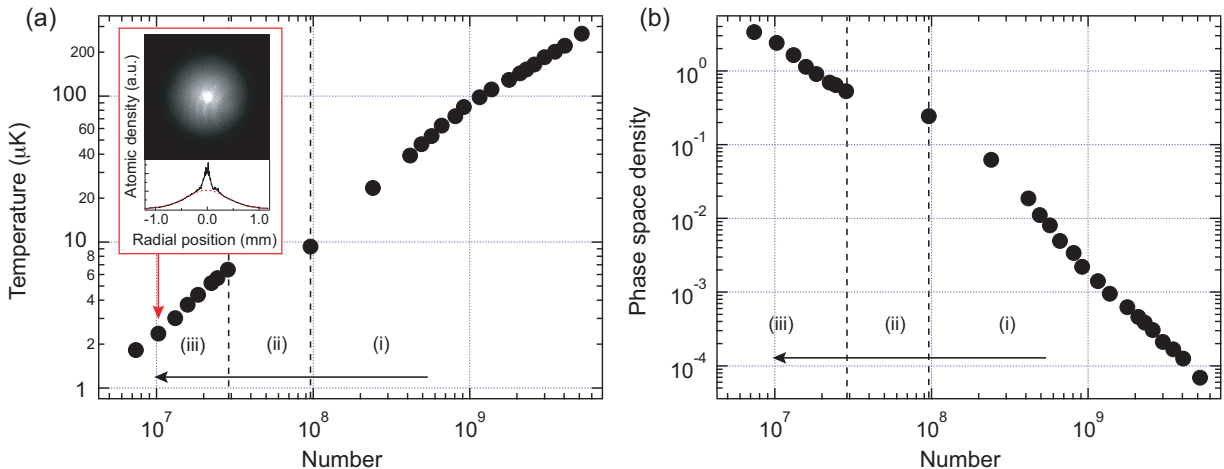}
\caption{Experimental path from the (i) to (iii) stage that produces large $^7$Li BECs. (i) Microwave evaporation in an optically plugged magnetic trap, (ii) transfer into a crossed dipole trap, and (iii) evaporation by trap tilting in the crossed optical trap. (a) Temperature $T$ vs atom number $N$. Inset: absorption image after 8~ms of expansion, indicating the onset of the BEC. (b) Peak phase-space density $D$ as a function of the atom number $N$. \label{EvapSum}}
\end{figure*}

Figure~5 shows the evaporation trajectory both in the magnetic and the optical dipole trap. The peak phase-space density $D=n\lambda_{dB}^3$ ($\lambda_{dB}=h/\sqrt{2\pi mk_BT}$ is the thermal de Broglie length) is increased five orders of magnitude by the evaporation cooling. The efficiency of evaporation, $\Gamma=-d(\ln D)/d(\ln N)$,  in each potential is 2.2 and 1.5, respectively. Since we are able to capture only $10^6$ number of atoms in the dipole trap at $100~\mu{}$K right after the molasses and the dark state pumping, the evaporation cooling in the quadrupole trap, which increases the peak phase-space density to a factor of $10^4$, is essential in obtaining large condensates. 

Cooling atoms with a trap tilt can be a new technique for studying non-equilibrium phenomena in the BEC phase transition of atomic gases~\cite{Weiler2008,Giacomo2013,Nir2015}. We search for the possibility of thermal quenching by increasing the field gradient to 60~G/cm in 65~ms. Here, optical decompression is not applied so that we can ignore the sloshing motions generated by a sudden change in trapping frequency during the rapid intensity ramp of evaporation. With the scheme, we can still make a pure condensate with $10^6$ atoms, but with a vortex, as shown in Fig.~4(c). The vortices appear more frequently as we increase the evaporation speed, suggesting the defects are nucleated by the Kibble-Zurek mechanism~\cite{Kibble1976,Zurek1985}. A detailed study of the vortex nucleation process and the vortex number scaling with quench time are worthy of future investigation.

\section{Conclusion}
We have described an experiment that rapidly produces large $^7$Li condensates in an optical dipole trap. Our method relies on the combination of gray molasses cooling on the $D_1$ transition line and two-stage evaporation cooling in a conservative trapping potential. The sub-Doppler cooling lowers the temperature of the atoms in the compressed MOT to $25~\mu$K, which might allow the all-optical production of $^7$Li BECs after directly loading the atoms into a deep optical potential~\cite{Burchianti2014,Salomon2014}. Runaway evaporation cooling is achieved in an optically plugged magnetic quadrupole trap, where the Majorana atom loss is highly suppressed by the plug beam. After subsequent evaporation in a crossed optical trap, we obtain a pure condensate of $\sim3\times10^6$ atoms. We adopt a trap-tilting scheme for a rapid evaporation in the optical trap, which can be useful for studying the Kibble-Zurek mechanism~\cite{Weiler2008,Giacomo2013,Nir2015} and universal dynamics in a far out-of-equilibrium state~\cite{berges2015,Erne2018} in an optical potential. 

\begin{acknowledgments}
The authors thank I.~Dimitrova, R.~Senaratne, and C. Gross for discussion about the experimental setup and H.~Jeong and W.~Noh for experimental help. This work was supported by National Research Foundation of Korea Grant No. 2017R1E1A1A01074161.
\end{acknowledgments}

\bibliography{Main}

\begin{thebibliography}{41}%
\makeatletter
\providecommand \@ifxundefined [1]{%
 \@ifx{#1\undefined}
}%
\providecommand \@ifnum [1]{%
 \ifnum #1\expandafter \@firstoftwo
 \else \expandafter \@secondoftwo
 \fi
}%
\providecommand \@ifx [1]{%
 \ifx #1\expandafter \@firstoftwo
 \else \expandafter \@secondoftwo
 \fi
}%
\providecommand \natexlab [1]{#1}%
\providecommand \enquote  [1]{``#1''}%
\providecommand \bibnamefont  [1]{#1}%
\providecommand \bibfnamefont [1]{#1}%
\providecommand \citenamefont [1]{#1}%
\providecommand \href@noop [0]{\@secondoftwo}%
\providecommand \href [0]{\begingroup \@sanitize@url \@href}%
\providecommand \@href[1]{\@@startlink{#1}\@@href}%
\providecommand \@@href[1]{\endgroup#1\@@endlink}%
\providecommand \@sanitize@url [0]{\catcode `\\12\catcode `\$12\catcode
  `\&12\catcode `\#12\catcode `\^12\catcode `\_12\catcode `\%12\relax}%
\providecommand \@@startlink[1]{}%
\providecommand \@@endlink[0]{}%
\providecommand \url  [0]{\begingroup\@sanitize@url \@url }%
\providecommand \@url [1]{\endgroup\@href {#1}{\urlprefix }}%
\providecommand \urlprefix  [0]{URL }%
\providecommand \Eprint [0]{\href }%
\providecommand \doibase [0]{http://dx.doi.org/}%
\providecommand \selectlanguage [0]{\@gobble}%
\providecommand \bibinfo  [0]{\@secondoftwo}%
\providecommand \bibfield  [0]{\@secondoftwo}%
\providecommand \translation [1]{[#1]}%
\providecommand \BibitemOpen [0]{}%
\providecommand \bibitemStop [0]{}%
\providecommand \bibitemNoStop [0]{.\EOS\space}%
\providecommand \EOS [0]{\spacefactor3000\relax}%
\providecommand \BibitemShut  [1]{\csname bibitem#1\endcsname}%
\let\auto@bib@innerbib\@empty
\bibitem [{\citenamefont {Bloch}\ \emph {et~al.}(2012)\citenamefont {Bloch},
  \citenamefont {Dalibard},\ and\ \citenamefont {Nascimb{\`e}ne}}]{Bloch2012}%
  \BibitemOpen
  \bibfield  {author} {\bibinfo {author} {\bibfnamefont {I.}~\bibnamefont
  {Bloch}}, \bibinfo {author} {\bibfnamefont {J.}~\bibnamefont {Dalibard}}, \
  and\ \bibinfo {author} {\bibfnamefont {S.}~\bibnamefont {Nascimb{\`e}ne}},\
  }\href {\doibase 10.1038/nphys2259} {\bibfield  {journal} {\bibinfo
  {journal} {Nat Phys}\ }\textbf {\bibinfo {volume} {8}},\ \bibinfo {pages}
  {267} (\bibinfo {year} {2012})}\BibitemShut {NoStop}%
\bibitem [{\citenamefont {Gross}\ and\ \citenamefont
  {Bloch}(2017)}]{Gross2017}%
  \BibitemOpen
  \bibfield  {author} {\bibinfo {author} {\bibfnamefont {C.}~\bibnamefont
  {Gross}}\ and\ \bibinfo {author} {\bibfnamefont {I.}~\bibnamefont {Bloch}},\
  }\href {\doibase 10.1126/science.aal3837} {\bibfield  {journal} {\bibinfo
  {journal} {Science}\ }\textbf {\bibinfo {volume} {357}},\ \bibinfo {pages}
  {995} (\bibinfo {year} {2017})}\BibitemShut {NoStop}%
\bibitem [{\citenamefont {Chin}\ \emph {et~al.}(2010)\citenamefont {Chin},
  \citenamefont {Grimm}, \citenamefont {Julienne},\ and\ \citenamefont
  {Tiesinga}}]{chin2010}%
  \BibitemOpen
  \bibfield  {author} {\bibinfo {author} {\bibfnamefont {C.}~\bibnamefont
  {Chin}}, \bibinfo {author} {\bibfnamefont {R.}~\bibnamefont {Grimm}},
  \bibinfo {author} {\bibfnamefont {P.}~\bibnamefont {Julienne}}, \ and\
  \bibinfo {author} {\bibfnamefont {E.}~\bibnamefont {Tiesinga}},\ }\href
  {\doibase 10.1103/RevModPhys.82.1225} {\bibfield  {journal} {\bibinfo
  {journal} {Rev. Mod. Phys.}\ }\textbf {\bibinfo {volume} {82}},\ \bibinfo
  {pages} {1225} (\bibinfo {year} {2010})}\BibitemShut {NoStop}%
\bibitem [{\citenamefont {Chevy}\ and\ \citenamefont
  {Salomon}(2016)}]{Chevy2016}%
  \BibitemOpen
  \bibfield  {author} {\bibinfo {author} {\bibfnamefont {F.}~\bibnamefont
  {Chevy}}\ and\ \bibinfo {author} {\bibfnamefont {C.}~\bibnamefont
  {Salomon}},\ }\href {\doibase 10.1088/0953-4075/49/19/192001} {\bibfield
  {journal} {\bibinfo  {journal} {J. Phys. B: At. Mol. Opt. Phys.}\ }\textbf
  {\bibinfo {volume} {49}},\ \bibinfo {pages} {192001} (\bibinfo {year}
  {2016})}\BibitemShut {NoStop}%
\bibitem [{\citenamefont {McDonald}\ \emph {et~al.}(2014)\citenamefont
  {McDonald}, \citenamefont {Kuhn}, \citenamefont {Hardman}, \citenamefont
  {Bennetts}, \citenamefont {Everitt}, \citenamefont {Altin}, \citenamefont
  {Debs}, \citenamefont {Close},\ and\ \citenamefont {Robins}}]{McDonald2014}%
  \BibitemOpen
  \bibfield  {author} {\bibinfo {author} {\bibfnamefont {G.~D.}\ \bibnamefont
  {McDonald}}, \bibinfo {author} {\bibfnamefont {C.~C.~N.}\ \bibnamefont
  {Kuhn}}, \bibinfo {author} {\bibfnamefont {K.~S.}\ \bibnamefont {Hardman}},
  \bibinfo {author} {\bibfnamefont {S.}~\bibnamefont {Bennetts}}, \bibinfo
  {author} {\bibfnamefont {P.~J.}\ \bibnamefont {Everitt}}, \bibinfo {author}
  {\bibfnamefont {P.~A.}\ \bibnamefont {Altin}}, \bibinfo {author}
  {\bibfnamefont {J.~E.}\ \bibnamefont {Debs}}, \bibinfo {author}
  {\bibfnamefont {J.~D.}\ \bibnamefont {Close}}, \ and\ \bibinfo {author}
  {\bibfnamefont {N.~P.}\ \bibnamefont {Robins}},\ }\href {\doibase
  10.1103/PhysRevLett.113.013002} {\bibfield  {journal} {\bibinfo  {journal}
  {Phys. Rev. Lett.}\ }\textbf {\bibinfo {volume} {113}},\ \bibinfo {pages}
  {013002} (\bibinfo {year} {2014})}\BibitemShut {NoStop}%
\bibitem [{\citenamefont {Ferrier-Barbut}\ \emph {et~al.}(2014)\citenamefont
  {Ferrier-Barbut}, \citenamefont {Delehaye}, \citenamefont {Laurent},
  \citenamefont {Grier}, \citenamefont {Pierce}, \citenamefont {Rem},
  \citenamefont {Chevy},\ and\ \citenamefont {Salomon}}]{Ferrier2014}%
  \BibitemOpen
  \bibfield  {author} {\bibinfo {author} {\bibfnamefont {I.}~\bibnamefont
  {Ferrier-Barbut}}, \bibinfo {author} {\bibfnamefont {M.}~\bibnamefont
  {Delehaye}}, \bibinfo {author} {\bibfnamefont {S.}~\bibnamefont {Laurent}},
  \bibinfo {author} {\bibfnamefont {A.~T.}\ \bibnamefont {Grier}}, \bibinfo
  {author} {\bibfnamefont {M.}~\bibnamefont {Pierce}}, \bibinfo {author}
  {\bibfnamefont {B.~S.}\ \bibnamefont {Rem}}, \bibinfo {author} {\bibfnamefont
  {F.}~\bibnamefont {Chevy}}, \ and\ \bibinfo {author} {\bibfnamefont
  {C.}~\bibnamefont {Salomon}},\ }\href {\doibase 10.1126/science.1255380}
  {\bibfield  {journal} {\bibinfo  {journal} {Science}\ }\textbf {\bibinfo
  {volume} {345}},\ \bibinfo {pages} {1035} (\bibinfo {year}
  {2014})}\BibitemShut {NoStop}%
\bibitem [{\citenamefont {Kuklov}\ and\ \citenamefont
  {Svistunov}(2003)}]{kuklov2003}%
  \BibitemOpen
  \bibfield  {author} {\bibinfo {author} {\bibfnamefont {A.~B.}\ \bibnamefont
  {Kuklov}}\ and\ \bibinfo {author} {\bibfnamefont {B.~V.}\ \bibnamefont
  {Svistunov}},\ }\href {\doibase 10.1103/PhysRevLett.90.100401} {\bibfield
  {journal} {\bibinfo  {journal} {Phys. Rev. Lett.}\ }\textbf {\bibinfo
  {volume} {90}},\ \bibinfo {pages} {100401} (\bibinfo {year}
  {2003})}\BibitemShut {NoStop}%
\bibitem [{\citenamefont {Duan}\ \emph {et~al.}(2003)\citenamefont {Duan},
  \citenamefont {Demler},\ and\ \citenamefont {Lukin}}]{Duan2003}%
  \BibitemOpen
  \bibfield  {author} {\bibinfo {author} {\bibfnamefont {L.-M.}\ \bibnamefont
  {Duan}}, \bibinfo {author} {\bibfnamefont {E.}~\bibnamefont {Demler}}, \ and\
  \bibinfo {author} {\bibfnamefont {M.~D.}\ \bibnamefont {Lukin}},\ }\href
  {\doibase 10.1103/PhysRevLett.91.090402} {\bibfield  {journal} {\bibinfo
  {journal} {Phys. Rev. Lett.}\ }\textbf {\bibinfo {volume} {91}},\ \bibinfo
  {pages} {090402} (\bibinfo {year} {2003})}\BibitemShut {NoStop}%
\bibitem [{\citenamefont {Lewenstein}\ \emph {et~al.}(2004)\citenamefont
  {Lewenstein}, \citenamefont {Santos}, \citenamefont {Baranov},\ and\
  \citenamefont {Fehrmann}}]{Lewenstein2004}%
  \BibitemOpen
  \bibfield  {author} {\bibinfo {author} {\bibfnamefont {M.}~\bibnamefont
  {Lewenstein}}, \bibinfo {author} {\bibfnamefont {L.}~\bibnamefont {Santos}},
  \bibinfo {author} {\bibfnamefont {M.~A.}\ \bibnamefont {Baranov}}, \ and\
  \bibinfo {author} {\bibfnamefont {H.}~\bibnamefont {Fehrmann}},\ }\href
  {\doibase 10.1103/PhysRevLett.92.050401} {\bibfield  {journal} {\bibinfo
  {journal} {Phys. Rev. Lett.}\ }\textbf {\bibinfo {volume} {92}},\ \bibinfo
  {pages} {050401} (\bibinfo {year} {2004})}\BibitemShut {NoStop}%
\bibitem [{\citenamefont {Bradley}\ \emph {et~al.}(1997)\citenamefont
  {Bradley}, \citenamefont {Sackett},\ and\ \citenamefont
  {Hulet}}]{Bradley1997}%
  \BibitemOpen
  \bibfield  {author} {\bibinfo {author} {\bibfnamefont {C.~C.}\ \bibnamefont
  {Bradley}}, \bibinfo {author} {\bibfnamefont {C.~A.}\ \bibnamefont
  {Sackett}}, \ and\ \bibinfo {author} {\bibfnamefont {R.~G.}\ \bibnamefont
  {Hulet}},\ }\href {\doibase 10.1103/PhysRevLett.78.985} {\bibfield  {journal}
  {\bibinfo  {journal} {Phys. Rev. Lett.}\ }\textbf {\bibinfo {volume} {78}},\
  \bibinfo {pages} {985} (\bibinfo {year} {1997})}\BibitemShut {NoStop}%
\bibitem [{\citenamefont {Gross}\ and\ \citenamefont
  {Khaykovich}(2008)}]{Gross2008}%
  \BibitemOpen
  \bibfield  {author} {\bibinfo {author} {\bibfnamefont {N.}~\bibnamefont
  {Gross}}\ and\ \bibinfo {author} {\bibfnamefont {L.}~\bibnamefont
  {Khaykovich}},\ }\href {\doibase 10.1103/PhysRevA.77.023604} {\bibfield
  {journal} {\bibinfo  {journal} {Phys. Rev. A}\ }\textbf {\bibinfo {volume}
  {77}},\ \bibinfo {pages} {023604} (\bibinfo {year} {2008})}\BibitemShut
  {NoStop}%
\bibitem [{\citenamefont {Schreck}\ \emph {et~al.}(2001)\citenamefont
  {Schreck}, \citenamefont {Khaykovich}, \citenamefont {Corwin}, \citenamefont
  {Ferrari}, \citenamefont {Bourdel}, \citenamefont {Cubizolles},\ and\
  \citenamefont {Salomon}}]{Schreck2001}%
  \BibitemOpen
  \bibfield  {author} {\bibinfo {author} {\bibfnamefont {F.}~\bibnamefont
  {Schreck}}, \bibinfo {author} {\bibfnamefont {L.}~\bibnamefont {Khaykovich}},
  \bibinfo {author} {\bibfnamefont {K.~L.}\ \bibnamefont {Corwin}}, \bibinfo
  {author} {\bibfnamefont {G.}~\bibnamefont {Ferrari}}, \bibinfo {author}
  {\bibfnamefont {T.}~\bibnamefont {Bourdel}}, \bibinfo {author} {\bibfnamefont
  {J.}~\bibnamefont {Cubizolles}}, \ and\ \bibinfo {author} {\bibfnamefont
  {C.}~\bibnamefont {Salomon}},\ }\href {\doibase
  10.1103/PhysRevLett.87.080403} {\bibfield  {journal} {\bibinfo  {journal}
  {Phys. Rev. Lett.}\ }\textbf {\bibinfo {volume} {87}},\ \bibinfo {pages}
  {080403} (\bibinfo {year} {2001})}\BibitemShut {NoStop}%
\bibitem [{\citenamefont {Ikemachi}\ \emph {et~al.}(2017)\citenamefont
  {Ikemachi}, \citenamefont {Ito}, \citenamefont {Aratake}, \citenamefont
  {Chen}, \citenamefont {Koashi}, \citenamefont {Kuwata-Gonokami},\ and\
  \citenamefont {Horikoshi}}]{Ikemachi2017}%
  \BibitemOpen
  \bibfield  {author} {\bibinfo {author} {\bibfnamefont {T.}~\bibnamefont
  {Ikemachi}}, \bibinfo {author} {\bibfnamefont {A.}~\bibnamefont {Ito}},
  \bibinfo {author} {\bibfnamefont {Y.}~\bibnamefont {Aratake}}, \bibinfo
  {author} {\bibfnamefont {Y.}~\bibnamefont {Chen}}, \bibinfo {author}
  {\bibfnamefont {M.}~\bibnamefont {Koashi}}, \bibinfo {author} {\bibfnamefont
  {M.}~\bibnamefont {Kuwata-Gonokami}}, \ and\ \bibinfo {author} {\bibfnamefont
  {M.}~\bibnamefont {Horikoshi}},\ }\href {\doibase
  10.1088/1361-6455/50/1/01lt01} {\bibfield  {journal} {\bibinfo  {journal} {J.
  Phys. B: At. Mol. Opt. Phys.}\ }\textbf {\bibinfo {volume} {50}},\ \bibinfo
  {pages} {01LT01} (\bibinfo {year} {2017})}\BibitemShut {NoStop}%
\bibitem [{\citenamefont {Grynberg}\ and\ \citenamefont
  {Courtois}(1994)}]{Grynberg1994}%
  \BibitemOpen
  \bibfield  {author} {\bibinfo {author} {\bibfnamefont {G.}~\bibnamefont
  {Grynberg}}\ and\ \bibinfo {author} {\bibfnamefont {J.-Y.}\ \bibnamefont
  {Courtois}},\ }\href {http://stacks.iop.org/0295-5075/27/i=1/a=008}
  {\bibfield  {journal} {\bibinfo  {journal} {EPL (Europhysics Letters)}\
  }\textbf {\bibinfo {volume} {27}},\ \bibinfo {pages} {41} (\bibinfo {year}
  {1994})}\BibitemShut {NoStop}%
\bibitem [{\citenamefont {Boiron}\ \emph {et~al.}(1995)\citenamefont {Boiron},
  \citenamefont {Trich\'e}, \citenamefont {Meacher}, \citenamefont {Verkerk},\
  and\ \citenamefont {Grynberg}}]{Boiron1995}%
  \BibitemOpen
  \bibfield  {author} {\bibinfo {author} {\bibfnamefont {D.}~\bibnamefont
  {Boiron}}, \bibinfo {author} {\bibfnamefont {C.}~\bibnamefont {Trich\'e}},
  \bibinfo {author} {\bibfnamefont {D.~R.}\ \bibnamefont {Meacher}}, \bibinfo
  {author} {\bibfnamefont {P.}~\bibnamefont {Verkerk}}, \ and\ \bibinfo
  {author} {\bibfnamefont {G.}~\bibnamefont {Grynberg}},\ }\href {\doibase
  10.1103/PhysRevA.52.R3425} {\bibfield  {journal} {\bibinfo  {journal} {Phys.
  Rev. A}\ }\textbf {\bibinfo {volume} {52}},\ \bibinfo {pages} {R3425}
  (\bibinfo {year} {1995})}\BibitemShut {NoStop}%
\bibitem [{\citenamefont {Esslinger}\ \emph {et~al.}(1996)\citenamefont
  {Esslinger}, \citenamefont {Sander}, \citenamefont {Hemmerich}, \citenamefont
  {H\"{a}nsch}, \citenamefont {Ritsch},\ and\ \citenamefont
  {Weidem\"{u}ller}}]{Esslinger1996}%
  \BibitemOpen
  \bibfield  {author} {\bibinfo {author} {\bibfnamefont {T.}~\bibnamefont
  {Esslinger}}, \bibinfo {author} {\bibfnamefont {F.}~\bibnamefont {Sander}},
  \bibinfo {author} {\bibfnamefont {A.}~\bibnamefont {Hemmerich}}, \bibinfo
  {author} {\bibfnamefont {T.~W.}\ \bibnamefont {H\"{a}nsch}}, \bibinfo
  {author} {\bibfnamefont {H.}~\bibnamefont {Ritsch}}, \ and\ \bibinfo {author}
  {\bibfnamefont {M.}~\bibnamefont {Weidem\"{u}ller}},\ }\href {\doibase
  10.1364/OL.21.000991} {\bibfield  {journal} {\bibinfo  {journal} {Opt.
  Lett.}\ }\textbf {\bibinfo {volume} {21}},\ \bibinfo {pages} {991} (\bibinfo
  {year} {1996})}\BibitemShut {NoStop}%
\bibitem [{\citenamefont {Boiron}\ \emph {et~al.}(1996)\citenamefont {Boiron},
  \citenamefont {Michaud}, \citenamefont {Lemonde}, \citenamefont {Castin},
  \citenamefont {Salomon}, \citenamefont {Weyers}, \citenamefont {Szymaniec},
  \citenamefont {Cognet},\ and\ \citenamefont {Clairon}}]{Boiron1996}%
  \BibitemOpen
  \bibfield  {author} {\bibinfo {author} {\bibfnamefont {D.}~\bibnamefont
  {Boiron}}, \bibinfo {author} {\bibfnamefont {A.}~\bibnamefont {Michaud}},
  \bibinfo {author} {\bibfnamefont {P.}~\bibnamefont {Lemonde}}, \bibinfo
  {author} {\bibfnamefont {Y.}~\bibnamefont {Castin}}, \bibinfo {author}
  {\bibfnamefont {C.}~\bibnamefont {Salomon}}, \bibinfo {author} {\bibfnamefont
  {S.}~\bibnamefont {Weyers}}, \bibinfo {author} {\bibfnamefont
  {K.}~\bibnamefont {Szymaniec}}, \bibinfo {author} {\bibfnamefont
  {L.}~\bibnamefont {Cognet}}, \ and\ \bibinfo {author} {\bibfnamefont
  {A.}~\bibnamefont {Clairon}},\ }\href {\doibase 10.1103/PhysRevA.53.R3734}
  {\bibfield  {journal} {\bibinfo  {journal} {Phys. Rev. A}\ }\textbf {\bibinfo
  {volume} {53}},\ \bibinfo {pages} {R3734} (\bibinfo {year}
  {1996})}\BibitemShut {NoStop}%
\bibitem [{\citenamefont {Fernandes}\ \emph {et~al.}(2012)\citenamefont
  {Fernandes}, \citenamefont {Sievers}, \citenamefont {Kretzschmar},
  \citenamefont {Wu}, \citenamefont {Salomon},\ and\ \citenamefont
  {Chevy}}]{Fernandes2012}%
  \BibitemOpen
  \bibfield  {author} {\bibinfo {author} {\bibfnamefont {R.~D.}\ \bibnamefont
  {Fernandes}}, \bibinfo {author} {\bibfnamefont {F.}~\bibnamefont {Sievers}},
  \bibinfo {author} {\bibfnamefont {N.}~\bibnamefont {Kretzschmar}}, \bibinfo
  {author} {\bibfnamefont {S.}~\bibnamefont {Wu}}, \bibinfo {author}
  {\bibfnamefont {C.}~\bibnamefont {Salomon}}, \ and\ \bibinfo {author}
  {\bibfnamefont {F.}~\bibnamefont {Chevy}},\ }\href {\doibase
  10.1209/0295-5075/100/63001} {\bibfield  {journal} {\bibinfo  {journal} {Epl
  Europhys Lett}\ }\textbf {\bibinfo {volume} {100}},\ \bibinfo {pages} {63001}
  (\bibinfo {year} {2012})}\BibitemShut {NoStop}%
\bibitem [{\citenamefont {Salomon}\ \emph {et~al.}(2013)\citenamefont
  {Salomon}, \citenamefont {Fouch{\'e}}, \citenamefont {Wang}, \citenamefont
  {Aspect}, \citenamefont {Bouyer},\ and\ \citenamefont
  {Bourdel}}]{Salomon2013}%
  \BibitemOpen
  \bibfield  {author} {\bibinfo {author} {\bibfnamefont {G.}~\bibnamefont
  {Salomon}}, \bibinfo {author} {\bibfnamefont {L.}~\bibnamefont {Fouch{\'e}}},
  \bibinfo {author} {\bibfnamefont {P.}~\bibnamefont {Wang}}, \bibinfo {author}
  {\bibfnamefont {A.}~\bibnamefont {Aspect}}, \bibinfo {author} {\bibfnamefont
  {P.}~\bibnamefont {Bouyer}}, \ and\ \bibinfo {author} {\bibfnamefont
  {T.}~\bibnamefont {Bourdel}},\ }\href {\doibase 10.1209/0295-5075/104/63002}
  {\bibfield  {journal} {\bibinfo  {journal} {Epl Europhys Lett}\ }\textbf
  {\bibinfo {volume} {104}},\ \bibinfo {pages} {63002} (\bibinfo {year}
  {2013})}\BibitemShut {NoStop}%
\bibitem [{\citenamefont {Grier}\ \emph {et~al.}(2013)\citenamefont {Grier},
  \citenamefont {Ferrier-Barbut}, \citenamefont {Rem}, \citenamefont
  {Delehaye}, \citenamefont {Khaykovich}, \citenamefont {Chevy},\ and\
  \citenamefont {Salomon}}]{grier2013lambda}%
  \BibitemOpen
  \bibfield  {author} {\bibinfo {author} {\bibfnamefont {A.~T.}\ \bibnamefont
  {Grier}}, \bibinfo {author} {\bibfnamefont {I.}~\bibnamefont
  {Ferrier-Barbut}}, \bibinfo {author} {\bibfnamefont {B.~S.}\ \bibnamefont
  {Rem}}, \bibinfo {author} {\bibfnamefont {M.}~\bibnamefont {Delehaye}},
  \bibinfo {author} {\bibfnamefont {L.}~\bibnamefont {Khaykovich}}, \bibinfo
  {author} {\bibfnamefont {F.}~\bibnamefont {Chevy}}, \ and\ \bibinfo {author}
  {\bibfnamefont {C.}~\bibnamefont {Salomon}},\ }\href {\doibase
  10.1103/PhysRevA.87.063411} {\bibfield  {journal} {\bibinfo  {journal} {Phys.
  Rev. A}\ }\textbf {\bibinfo {volume} {87}},\ \bibinfo {pages} {063411}
  (\bibinfo {year} {2013})}\BibitemShut {NoStop}%
\bibitem [{\citenamefont {Burchianti}\ \emph {et~al.}(2014)\citenamefont
  {Burchianti}, \citenamefont {Valtolina}, \citenamefont {Seman}, \citenamefont
  {Pace}, \citenamefont {Pas}, \citenamefont {Inguscio}, \citenamefont
  {Zaccanti},\ and\ \citenamefont {Roati}}]{Burchianti2014}%
  \BibitemOpen
  \bibfield  {author} {\bibinfo {author} {\bibfnamefont {A.}~\bibnamefont
  {Burchianti}}, \bibinfo {author} {\bibfnamefont {G.}~\bibnamefont
  {Valtolina}}, \bibinfo {author} {\bibfnamefont {J.}~\bibnamefont {Seman}},
  \bibinfo {author} {\bibfnamefont {E.}~\bibnamefont {Pace}}, \bibinfo {author}
  {\bibfnamefont {D.~M.}\ \bibnamefont {Pas}}, \bibinfo {author} {\bibfnamefont
  {M.}~\bibnamefont {Inguscio}}, \bibinfo {author} {\bibfnamefont
  {M.}~\bibnamefont {Zaccanti}}, \ and\ \bibinfo {author} {\bibfnamefont
  {G.}~\bibnamefont {Roati}},\ }\href {\doibase 10.1103/PhysRevA.90.043408}
  {\bibfield  {journal} {\bibinfo  {journal} {Phys Rev A}\ }\textbf {\bibinfo
  {volume} {90}},\ \bibinfo {pages} {043408} (\bibinfo {year}
  {2014})}\BibitemShut {NoStop}%
\bibitem [{\citenamefont {Colzi}\ \emph {et~al.}(2016)\citenamefont {Colzi},
  \citenamefont {Durastante}, \citenamefont {Fava}, \citenamefont {Serafini},
  \citenamefont {Lamporesi},\ and\ \citenamefont {Ferrari}}]{colzi2016sub}%
  \BibitemOpen
  \bibfield  {author} {\bibinfo {author} {\bibfnamefont {G.}~\bibnamefont
  {Colzi}}, \bibinfo {author} {\bibfnamefont {G.}~\bibnamefont {Durastante}},
  \bibinfo {author} {\bibfnamefont {E.}~\bibnamefont {Fava}}, \bibinfo {author}
  {\bibfnamefont {S.}~\bibnamefont {Serafini}}, \bibinfo {author}
  {\bibfnamefont {G.}~\bibnamefont {Lamporesi}}, \ and\ \bibinfo {author}
  {\bibfnamefont {G.}~\bibnamefont {Ferrari}},\ }\href@noop {} {\bibfield
  {journal} {\bibinfo  {journal} {Physical Review A}\ }\textbf {\bibinfo
  {volume} {93}},\ \bibinfo {pages} {023421} (\bibinfo {year}
  {2016})}\BibitemShut {NoStop}%
\bibitem [{\citenamefont {Dimitrova}\ \emph {et~al.}(2017)\citenamefont
  {Dimitrova}, \citenamefont {Lunden}, \citenamefont {Amato-Grill},
  \citenamefont {Jepsen}, \citenamefont {Yu}, \citenamefont {Messer},
  \citenamefont {Rigaldo}, \citenamefont {Puentes}, \citenamefont {Weld},\ and\
  \citenamefont {Ketterle}}]{Dimitrova2017}%
  \BibitemOpen
  \bibfield  {author} {\bibinfo {author} {\bibfnamefont {I.}~\bibnamefont
  {Dimitrova}}, \bibinfo {author} {\bibfnamefont {W.}~\bibnamefont {Lunden}},
  \bibinfo {author} {\bibfnamefont {J.}~\bibnamefont {Amato-Grill}}, \bibinfo
  {author} {\bibfnamefont {N.}~\bibnamefont {Jepsen}}, \bibinfo {author}
  {\bibfnamefont {Y.}~\bibnamefont {Yu}}, \bibinfo {author} {\bibfnamefont
  {M.}~\bibnamefont {Messer}}, \bibinfo {author} {\bibfnamefont
  {T.}~\bibnamefont {Rigaldo}}, \bibinfo {author} {\bibfnamefont
  {G.}~\bibnamefont {Puentes}}, \bibinfo {author} {\bibfnamefont
  {D.}~\bibnamefont {Weld}}, \ and\ \bibinfo {author} {\bibfnamefont
  {W.}~\bibnamefont {Ketterle}},\ }\href {\doibase 10.1103/PhysRevA.96.051603}
  {\bibfield  {journal} {\bibinfo  {journal} {Phys. Rev. A}\ }\textbf {\bibinfo
  {volume} {96}},\ \bibinfo {pages} {051603} (\bibinfo {year}
  {2017})}\BibitemShut {NoStop}%
\bibitem [{\citenamefont {Geiger}\ \emph {et~al.}(2018)\citenamefont {Geiger},
  \citenamefont {Fujiwara}, \citenamefont {Singh}, \citenamefont {Senaratne},
  \citenamefont {Rajagopal}, \citenamefont {Lipatov}, \citenamefont
  {Shimasaki}, \citenamefont {Driben}, \citenamefont {Konotop}, \citenamefont
  {Meier},\ and\ \citenamefont {Weld}}]{Geiger2018}%
  \BibitemOpen
  \bibfield  {author} {\bibinfo {author} {\bibfnamefont {Z.~A.}\ \bibnamefont
  {Geiger}}, \bibinfo {author} {\bibfnamefont {K.~M.}\ \bibnamefont
  {Fujiwara}}, \bibinfo {author} {\bibfnamefont {K.}~\bibnamefont {Singh}},
  \bibinfo {author} {\bibfnamefont {R.}~\bibnamefont {Senaratne}}, \bibinfo
  {author} {\bibfnamefont {S.~V.}\ \bibnamefont {Rajagopal}}, \bibinfo {author}
  {\bibfnamefont {M.}~\bibnamefont {Lipatov}}, \bibinfo {author} {\bibfnamefont
  {T.}~\bibnamefont {Shimasaki}}, \bibinfo {author} {\bibfnamefont
  {R.}~\bibnamefont {Driben}}, \bibinfo {author} {\bibfnamefont {V.~V.}\
  \bibnamefont {Konotop}}, \bibinfo {author} {\bibfnamefont {T.}~\bibnamefont
  {Meier}}, \ and\ \bibinfo {author} {\bibfnamefont {D.~M.}\ \bibnamefont
  {Weld}},\ }\href {\doibase 10.1103/PhysRevLett.120.213201} {\bibfield
  {journal} {\bibinfo  {journal} {Phys. Rev. Lett.}\ }\textbf {\bibinfo
  {volume} {120}},\ \bibinfo {pages} {213201} (\bibinfo {year}
  {2018})}\BibitemShut {NoStop}%
\bibitem [{\citenamefont {Hung}\ \emph {et~al.}(2008)\citenamefont {Hung},
  \citenamefont {Zhang}, \citenamefont {Gemelke},\ and\ \citenamefont
  {Chin}}]{hung2008}%
  \BibitemOpen
  \bibfield  {author} {\bibinfo {author} {\bibfnamefont {C.-L.}\ \bibnamefont
  {Hung}}, \bibinfo {author} {\bibfnamefont {X.}~\bibnamefont {Zhang}},
  \bibinfo {author} {\bibfnamefont {N.}~\bibnamefont {Gemelke}}, \ and\
  \bibinfo {author} {\bibfnamefont {C.}~\bibnamefont {Chin}},\ }\href@noop {}
  {\bibfield  {journal} {\bibinfo  {journal} {Physical Review A}\ }\textbf
  {\bibinfo {volume} {78}},\ \bibinfo {pages} {011604} (\bibinfo {year}
  {2008})}\BibitemShut {NoStop}%
\bibitem [{\citenamefont {Kibble}(1976)}]{Kibble1976}%
  \BibitemOpen
  \bibfield  {author} {\bibinfo {author} {\bibfnamefont {T.~W.~B.}\
  \bibnamefont {Kibble}},\ }\href {http://stacks.iop.org/0305-4470/9/i=8/a=029}
  {\bibfield  {journal} {\bibinfo  {journal} {Journal of Physics A}\ }\textbf
  {\bibinfo {volume} {9}},\ \bibinfo {pages} {1387} (\bibinfo {year}
  {1976})}\BibitemShut {NoStop}%
\bibitem [{\citenamefont {Zurek}(1985)}]{Zurek1985}%
  \BibitemOpen
  \bibfield  {author} {\bibinfo {author} {\bibfnamefont {W.}~\bibnamefont
  {Zurek}},\ }\href {\doibase 10.1038/317505a0} {\bibfield  {journal} {\bibinfo
   {journal} {Nature}\ }\textbf {\bibinfo {volume} {317}},\ \bibinfo {pages}
  {505} (\bibinfo {year} {1985})}\BibitemShut {NoStop}%
\bibitem [{\citenamefont {Davis}\ \emph {et~al.}(1995)\citenamefont {Davis},
  \citenamefont {Mewes}, \citenamefont {Andrews}, \citenamefont {van Druten},
  \citenamefont {Durfee}, \citenamefont {Kurn},\ and\ \citenamefont
  {Ketterle}}]{Davis1995}%
  \BibitemOpen
  \bibfield  {author} {\bibinfo {author} {\bibfnamefont {K.~B.}\ \bibnamefont
  {Davis}}, \bibinfo {author} {\bibfnamefont {M.~O.}\ \bibnamefont {Mewes}},
  \bibinfo {author} {\bibfnamefont {M.~R.}\ \bibnamefont {Andrews}}, \bibinfo
  {author} {\bibfnamefont {N.~J.}\ \bibnamefont {van Druten}}, \bibinfo
  {author} {\bibfnamefont {D.~S.}\ \bibnamefont {Durfee}}, \bibinfo {author}
  {\bibfnamefont {D.~M.}\ \bibnamefont {Kurn}}, \ and\ \bibinfo {author}
  {\bibfnamefont {W.}~\bibnamefont {Ketterle}},\ }\href {\doibase
  10.1103/PhysRevLett.75.3969} {\bibfield  {journal} {\bibinfo  {journal}
  {Phys. Rev. Lett.}\ }\textbf {\bibinfo {volume} {75}},\ \bibinfo {pages}
  {3969} (\bibinfo {year} {1995})}\BibitemShut {NoStop}%
\bibitem [{\citenamefont {Aspect}\ \emph {et~al.}(1988)\citenamefont {Aspect},
  \citenamefont {Arimondo}, \citenamefont {Kaiser}, \citenamefont
  {Vansteenkiste},\ and\ \citenamefont {Cohen-Tannoudji}}]{Aspect1988}%
  \BibitemOpen
  \bibfield  {author} {\bibinfo {author} {\bibfnamefont {A.}~\bibnamefont
  {Aspect}}, \bibinfo {author} {\bibfnamefont {E.}~\bibnamefont {Arimondo}},
  \bibinfo {author} {\bibfnamefont {R.}~\bibnamefont {Kaiser}}, \bibinfo
  {author} {\bibfnamefont {N.}~\bibnamefont {Vansteenkiste}}, \ and\ \bibinfo
  {author} {\bibfnamefont {C.}~\bibnamefont {Cohen-Tannoudji}},\ }\href
  {\doibase 10.1103/PhysRevLett.61.826} {\bibfield  {journal} {\bibinfo
  {journal} {Phys. Rev. Lett.}\ }\textbf {\bibinfo {volume} {61}},\ \bibinfo
  {pages} {826} (\bibinfo {year} {1988})}\BibitemShut {NoStop}%
\bibitem [{\citenamefont {Truscott}\ \emph {et~al.}(2001)\citenamefont
  {Truscott}, \citenamefont {Strecker}, \citenamefont {McAlexander},
  \citenamefont {Partridge},\ and\ \citenamefont {Hulet}}]{Truscott2001}%
  \BibitemOpen
  \bibfield  {author} {\bibinfo {author} {\bibfnamefont {A.~G.}\ \bibnamefont
  {Truscott}}, \bibinfo {author} {\bibfnamefont {K.~E.}\ \bibnamefont
  {Strecker}}, \bibinfo {author} {\bibfnamefont {W.~I.}\ \bibnamefont
  {McAlexander}}, \bibinfo {author} {\bibfnamefont {G.~B.}\ \bibnamefont
  {Partridge}}, \ and\ \bibinfo {author} {\bibfnamefont {R.~G.}\ \bibnamefont
  {Hulet}},\ }\href {\doibase 10.1126/science.1059318} {\bibfield  {journal}
  {\bibinfo  {journal} {Science}\ }\textbf {\bibinfo {volume} {291}},\ \bibinfo
  {pages} {2570} (\bibinfo {year} {2001})}\BibitemShut {NoStop}%
\bibitem [{\citenamefont {Abraham}\ \emph {et~al.}(1997)\citenamefont
  {Abraham}, \citenamefont {McAlexander}, \citenamefont {Gerton}, \citenamefont
  {Hulet}, \citenamefont {C\^ot\'e},\ and\ \citenamefont
  {Dalgarno}}]{Abraham1997}%
  \BibitemOpen
  \bibfield  {author} {\bibinfo {author} {\bibfnamefont {E.~R.~I.}\
  \bibnamefont {Abraham}}, \bibinfo {author} {\bibfnamefont {W.~I.}\
  \bibnamefont {McAlexander}}, \bibinfo {author} {\bibfnamefont {J.~M.}\
  \bibnamefont {Gerton}}, \bibinfo {author} {\bibfnamefont {R.~G.}\
  \bibnamefont {Hulet}}, \bibinfo {author} {\bibfnamefont {R.}~\bibnamefont
  {C\^ot\'e}}, \ and\ \bibinfo {author} {\bibfnamefont {A.}~\bibnamefont
  {Dalgarno}},\ }\href {\doibase 10.1103/PhysRevA.55.R3299} {\bibfield
  {journal} {\bibinfo  {journal} {Phys. Rev. A}\ }\textbf {\bibinfo {volume}
  {55}},\ \bibinfo {pages} {R3299} (\bibinfo {year} {1997})}\BibitemShut
  {NoStop}%
\bibitem [{\citenamefont {Gerton}\ \emph {et~al.}(1999)\citenamefont {Gerton},
  \citenamefont {Sackett}, \citenamefont {Frew},\ and\ \citenamefont
  {Hulet}}]{Gerton1999}%
  \BibitemOpen
  \bibfield  {author} {\bibinfo {author} {\bibfnamefont {J.~M.}\ \bibnamefont
  {Gerton}}, \bibinfo {author} {\bibfnamefont {C.~A.}\ \bibnamefont {Sackett}},
  \bibinfo {author} {\bibfnamefont {B.~J.}\ \bibnamefont {Frew}}, \ and\
  \bibinfo {author} {\bibfnamefont {R.~G.}\ \bibnamefont {Hulet}},\ }\href
  {\doibase 10.1103/PhysRevA.59.1514} {\bibfield  {journal} {\bibinfo
  {journal} {Phys. Rev. A}\ }\textbf {\bibinfo {volume} {59}},\ \bibinfo
  {pages} {1514} (\bibinfo {year} {1999})}\BibitemShut {NoStop}%
\bibitem [{\citenamefont {Petrich}\ \emph {et~al.}(1995)\citenamefont
  {Petrich}, \citenamefont {Anderson}, \citenamefont {Ensher},\ and\
  \citenamefont {Cornell}}]{Petrich1995}%
  \BibitemOpen
  \bibfield  {author} {\bibinfo {author} {\bibfnamefont {W.}~\bibnamefont
  {Petrich}}, \bibinfo {author} {\bibfnamefont {M.~H.}\ \bibnamefont
  {Anderson}}, \bibinfo {author} {\bibfnamefont {J.~R.}\ \bibnamefont
  {Ensher}}, \ and\ \bibinfo {author} {\bibfnamefont {E.~A.}\ \bibnamefont
  {Cornell}},\ }\href {\doibase 10.1103/PhysRevLett.74.3352} {\bibfield
  {journal} {\bibinfo  {journal} {Phys. Rev. Lett.}\ }\textbf {\bibinfo
  {volume} {74}},\ \bibinfo {pages} {3352} (\bibinfo {year}
  {1995})}\BibitemShut {NoStop}%
\bibitem [{\citenamefont {Heo}\ \emph {et~al.}(2011)\citenamefont {Heo},
  \citenamefont {Choi},\ and\ \citenamefont {Shin}}]{Heo2011}%
  \BibitemOpen
  \bibfield  {author} {\bibinfo {author} {\bibfnamefont {M.-S.}\ \bibnamefont
  {Heo}}, \bibinfo {author} {\bibfnamefont {J.-y.}\ \bibnamefont {Choi}}, \
  and\ \bibinfo {author} {\bibfnamefont {Y.-i.}\ \bibnamefont {Shin}},\ }\href
  {\doibase 10.1103/PhysRevA.83.013622} {\bibfield  {journal} {\bibinfo
  {journal} {Phys. Rev. A}\ }\textbf {\bibinfo {volume} {83}},\ \bibinfo
  {pages} {013622} (\bibinfo {year} {2011})}\BibitemShut {NoStop}%
\bibitem [{\citenamefont {Dubessy}\ \emph {et~al.}(2012)\citenamefont
  {Dubessy}, \citenamefont {Merloti}, \citenamefont {Longchambon},
  \citenamefont {Pottie}, \citenamefont {Liennard}, \citenamefont {Perrin},
  \citenamefont {Lorent},\ and\ \citenamefont {Perrin}}]{dubessy2012rubidium}%
  \BibitemOpen
  \bibfield  {author} {\bibinfo {author} {\bibfnamefont {R.}~\bibnamefont
  {Dubessy}}, \bibinfo {author} {\bibfnamefont {K.}~\bibnamefont {Merloti}},
  \bibinfo {author} {\bibfnamefont {L.}~\bibnamefont {Longchambon}}, \bibinfo
  {author} {\bibfnamefont {P.-E.}\ \bibnamefont {Pottie}}, \bibinfo {author}
  {\bibfnamefont {T.}~\bibnamefont {Liennard}}, \bibinfo {author}
  {\bibfnamefont {A.}~\bibnamefont {Perrin}}, \bibinfo {author} {\bibfnamefont
  {V.}~\bibnamefont {Lorent}}, \ and\ \bibinfo {author} {\bibfnamefont
  {H.}~\bibnamefont {Perrin}},\ }\href {\doibase 10.1103/PhysRevA.85.013643}
  {\bibfield  {journal} {\bibinfo  {journal} {Phys. Rev. A}\ }\textbf {\bibinfo
  {volume} {85}},\ \bibinfo {pages} {013643} (\bibinfo {year}
  {2012})}\BibitemShut {NoStop}%
\bibitem [{\citenamefont {Weiler}\ \emph {et~al.}(2008)\citenamefont {Weiler},
  \citenamefont {Neely}, \citenamefont {Scherer}, \citenamefont {Bradley},
  \citenamefont {Davis},\ and\ \citenamefont {Anderson}}]{Weiler2008}%
  \BibitemOpen
  \bibfield  {author} {\bibinfo {author} {\bibfnamefont {C.~N.}\ \bibnamefont
  {Weiler}}, \bibinfo {author} {\bibfnamefont {T.~W.}\ \bibnamefont {Neely}},
  \bibinfo {author} {\bibfnamefont {D.~R.}\ \bibnamefont {Scherer}}, \bibinfo
  {author} {\bibfnamefont {A.~S.}\ \bibnamefont {Bradley}}, \bibinfo {author}
  {\bibfnamefont {M.~J.}\ \bibnamefont {Davis}}, \ and\ \bibinfo {author}
  {\bibfnamefont {B.~P.}\ \bibnamefont {Anderson}},\ }\href
  {https://www.nature.com/articles/nature07334} {\bibfield  {journal} {\bibinfo
   {journal} {Nature}\ }\textbf {\bibinfo {volume} {455}},\ \bibinfo {pages}
  {948} (\bibinfo {year} {2008})}\BibitemShut {NoStop}%
\bibitem [{\citenamefont {Lamporesi}\ \emph {et~al.}(2013)\citenamefont
  {Lamporesi}, \citenamefont {Donadello}, \citenamefont {Serafini},
  \citenamefont {Dalfovo},\ and\ \citenamefont {Ferrari}}]{Giacomo2013}%
  \BibitemOpen
  \bibfield  {author} {\bibinfo {author} {\bibfnamefont {G.}~\bibnamefont
  {Lamporesi}}, \bibinfo {author} {\bibfnamefont {S.}~\bibnamefont
  {Donadello}}, \bibinfo {author} {\bibfnamefont {S.}~\bibnamefont {Serafini}},
  \bibinfo {author} {\bibfnamefont {F.}~\bibnamefont {Dalfovo}}, \ and\
  \bibinfo {author} {\bibfnamefont {G.}~\bibnamefont {Ferrari}},\ }\href
  {\doibase 10.1038/nphys2734} {\bibfield  {journal} {\bibinfo  {journal} {Nat
  Phys}\ }\textbf {\bibinfo {volume} {9}},\ \bibinfo {pages} {656} (\bibinfo
  {year} {2013})}\BibitemShut {NoStop}%
\bibitem [{\citenamefont {Navon}\ \emph {et~al.}(2015)\citenamefont {Navon},
  \citenamefont {Gaunt}, \citenamefont {Smith},\ and\ \citenamefont
  {Hadzibabic}}]{Nir2015}%
  \BibitemOpen
  \bibfield  {author} {\bibinfo {author} {\bibfnamefont {N.}~\bibnamefont
  {Navon}}, \bibinfo {author} {\bibfnamefont {A.~L.}\ \bibnamefont {Gaunt}},
  \bibinfo {author} {\bibfnamefont {R.~P.}\ \bibnamefont {Smith}}, \ and\
  \bibinfo {author} {\bibfnamefont {Z.}~\bibnamefont {Hadzibabic}},\ }\href
  {\doibase 10.1126/science.1258676} {\bibfield  {journal} {\bibinfo  {journal}
  {Science}\ }\textbf {\bibinfo {volume} {347}},\ \bibinfo {pages} {167}
  (\bibinfo {year} {2015})}\BibitemShut {NoStop}%
\bibitem [{\citenamefont {Salomon}\ \emph {et~al.}(2014)\citenamefont
  {Salomon}, \citenamefont {Fouch\'e}, \citenamefont {Lepoutre}, \citenamefont
  {Aspect},\ and\ \citenamefont {Bourdel}}]{Salomon2014}%
  \BibitemOpen
  \bibfield  {author} {\bibinfo {author} {\bibfnamefont {G.}~\bibnamefont
  {Salomon}}, \bibinfo {author} {\bibfnamefont {L.}~\bibnamefont {Fouch\'e}},
  \bibinfo {author} {\bibfnamefont {S.}~\bibnamefont {Lepoutre}}, \bibinfo
  {author} {\bibfnamefont {A.}~\bibnamefont {Aspect}}, \ and\ \bibinfo {author}
  {\bibfnamefont {T.}~\bibnamefont {Bourdel}},\ }\href {\doibase
  10.1103/PhysRevA.90.033405} {\bibfield  {journal} {\bibinfo  {journal} {Phys.
  Rev. A}\ }\textbf {\bibinfo {volume} {90}},\ \bibinfo {pages} {033405}
  (\bibinfo {year} {2014})}\BibitemShut {NoStop}%
\bibitem [{\citenamefont {Berges}\ \emph {et~al.}(2015)\citenamefont {Berges},
  \citenamefont {Boguslavski}, \citenamefont {Schlichting},\ and\ \citenamefont
  {Venugopalan}}]{berges2015}%
  \BibitemOpen
  \bibfield  {author} {\bibinfo {author} {\bibfnamefont {J.}~\bibnamefont
  {Berges}}, \bibinfo {author} {\bibfnamefont {K.}~\bibnamefont {Boguslavski}},
  \bibinfo {author} {\bibfnamefont {S.}~\bibnamefont {Schlichting}}, \ and\
  \bibinfo {author} {\bibfnamefont {R.}~\bibnamefont {Venugopalan}},\ }\href
  {\doibase 10.1103/PhysRevLett.114.061601} {\bibfield  {journal} {\bibinfo
  {journal} {Phys. Rev. Lett.}\ }\textbf {\bibinfo {volume} {114}},\ \bibinfo
  {pages} {061601} (\bibinfo {year} {2015})}\BibitemShut {NoStop}%
\bibitem [{\citenamefont {Erne}\ \emph {et~al.}(2018)\citenamefont {Erne},
  \citenamefont {B{\"u}cker}, \citenamefont {Gasenzer}, \citenamefont
  {Berges},\ and\ \citenamefont {Schmiedmayer}}]{Erne2018}%
  \BibitemOpen
  \bibfield  {author} {\bibinfo {author} {\bibfnamefont {S.}~\bibnamefont
  {Erne}}, \bibinfo {author} {\bibfnamefont {R.}~\bibnamefont {B{\"u}cker}},
  \bibinfo {author} {\bibfnamefont {T.}~\bibnamefont {Gasenzer}}, \bibinfo
  {author} {\bibfnamefont {J.}~\bibnamefont {Berges}}, \ and\ \bibinfo {author}
  {\bibfnamefont {J.}~\bibnamefont {Schmiedmayer}},\ }\href {\doibase
  10.1038/s41586-018-0667-0} {\bibfield  {journal} {\bibinfo  {journal}
  {Nature}\ }\textbf {\bibinfo {volume} {563}},\ \bibinfo {pages} {225}
  (\bibinfo {year} {2018})}\BibitemShut {NoStop}%
\end{thebibliography}%

\end{document}